\documentstyle[12pt,a4]{article}
\begin{document}
\thispagestyle{empty}
\begin{center}
\LARGE
CHAOS IN NEURAL NETWORKS WITH A NONMONOTONIC TRANSFER FUNCTION
~\\
~\\
\vspace{1.cm}
\normalsize
D. Caroppo, M. Mannarelli, G. Nardulli and S. Stramaglia\\
\vspace{0.5cm}
{\it Dipartimento Interateneo di Fisica} and\\
{\it Istituto Nazionale di Fisica Nucleare, Sezione di Bari\\
via Amendola 173, 70126 Bari, Italy}\\
~\\
~\\
\end{center}
\vspace{1.cm}
\begin{abstract}
Time evolution of diluted neural networks with a nonmonotonic transfer function 
is 
analitically described by flow equations for macroscopic variables. The
macroscopic dynamics shows a rich variety of
behaviours: fixed-point, periodicity and chaos. 
We examine in detail the structure of the
strange attractor and in particular we study the main features of the stable 
and unstable 
manifolds, the hyperbolicity of the attractor and the existence of homoclinic 
intersections. 
We also discuss the problem of the robustness of the chaos and we prove that in 
the present model chaotic behaviour is fragile (chaotic regions are densely 
intercalated with periodicity windows), according to a recently discussed 
conjecture.
Finally we perform  an analysis of the microscopic behaviour and in
particular
we examine the occurrence of damage spreading by studying the time evolution of
two almost identical initial configurations. We show that for any choice of the 
parameters 
the two initial states remain microscopically distinct.  
\end{abstract}

\vspace{1.cm}
\vskip .5cm
\noindent
%Keywords: Neural Networks, Non-monotone transfer function, Complex dynamics.
\vskip 1.cm
\noindent

\newpage
\addtolength{\baselineskip}{\baselineskip}

\section{Introduction}
Since the pioneering work by Sompolinsky et al.~\cite{somp}, the
occurrence of oscillations and chaos has become a major field of
interest in the frame of neural networks~\cite{bo}. Neural networks with
symmetric synaptic connections have been object of extensive studies by methods 
closely related to those used in the theoretical description of the spin 
glasses~\cite{fish}, since they admit an energy function. Also asymmetric 
synapses have 
been studied and the presence of   chaotic dynamics  was examined, 
following \cite{somp}, 
in a number of subsequent papers (see e.g.~
\cite{tir,doyon,ces,molg,nn1}). The investigation of chaotic neural networks
is interesting not only from a theoretical point of view, but also for 
practical  
reasons, as their dynamical possibilities are richer and allow for a larger 
spectrum of 
engineering applications (see, e.g., Ref.~\cite{hir}). 
It is also worth stressing that the brain is a highly dynamic system. The rich 
temporal structure (oscillations) of neural processes has been studied in 
~\cite{cerv1,cerv2,cerv4,cerv5}; 
chaotic behaviour has been found out in the nervous 
system~\cite{cerv3}. Relying on these neurophysiological findings, the study of 
chaos in neural networks may be useful in the comprehension of cognitive 
processes in the brain~\cite{ska}.

Asymmetric synapses are not the only route to chaos in neural networks; 
another possibility is to use a nonmonotonic functional dependence for the 
activation function,
i.e. the transfer function that gives the state of the neuron as a function of
the post-synaptic potential. In recent papers \cite{bolle,domin} it has been 
shown that such a
nonmonotonic transfer function may lead to macroscopic chaos in
attractor  neural networks: chaos appears in a class of macroscopic trajectories
characterized by an  overlap with the initial configuration that never vanishes.
In other words, the network preserves a memory of the
initial configuration, but the macroscopic overlap does not converge to a
fixed value and oscillates giving rise to a chaotic time series. Also the case
of diluted networks with dynamical, adaptative synapses and
nonmonotonic neurons in presence of
a Hebbian learning mechanism has been studied,and
it has been found that the adaptation leads to reduction of dynamics~\cite{cs}.

In this paper we further analyze the dynamic behaviour of attractor
neural networks with nonmonotonic transfer function. In particular we
analyze a network by mean-field equations whose macroscopic dynamics can
be analitically calculated~\cite{de,noi}. The time evolution of the
macroscopic parameters describing the system is determined by a
two-dimensional map that exhibits chaotic behaviour and represents in our 
opinion a non 
trivial and interesting example of a 
non-linear dynamical system (for recent reviews see e.g. \cite{ott,arrow}). 
In the 
present 
work the following 
issues are considered: structure and hyperbolicity of the strange attractor,
Hausdorff dimension and Lyapunov exponents. These are typical analyses of the 
non linear dynamical behaviour that we perform in a neural-motivated 
two-dimensional map in order to achieve a better understanding of the dynamics 
of this class of neural networks.
We also analyze the problem of the fragility of chaos and we explicitly prove 
that the present model behaves in agreement with the
conjecture 
in~\cite{barr}, i.e. that periodicity windows are constructed around spine 
loci (one-dimensional manifolds in the two-dimensional parameter space of the 
model here considered). This is in our opinion an interesting confirmation of 
this conjecture that sheds light on the geometrical features of the periodicity 
windows to be found in the chaotic regions.
 Finally, we examine the 
microscopic behaviour 
underlying the mean field description: we consider  two
replicas of the system starting from slightly different initial
conditions and  we show that these two different configurations never become 
identical,
independently of their macroscopic behaviour.
This feature was already observed  in diluted networks with monotone
transfer function~\cite{de}; here we prove that such behaviour is also 
present in the 
case of nonmonotonic neurons. It follows that at microscopic level the
network dynamics is always to be considered chaotic, whereas from a macroscopic,
mean field point of view, a rich variety of behaviours can occurr: fixed-point,
periodicity and chaos. We note that a similar emergence of a macroscopic 
evolution in presence of microscopic chaos has been recently found in another 
framework, i.e. Chaotic Coupled Maps models, and it has been termed {\it 
nontrivial collective behaviour} (NTCB, see~\cite{ren} and references therein).

The paper is organized as follows: in the next section the model is
described and the flow equations for macroscopic parameters are
reported and analyzed. In section 3 we study the time evolution of the
distance between two replicas of the network. In section 4 we present our
conclusions.
  
\section{The model: analysis of flow equations.}
We consider the model of Ref.~\cite{cs}, i.e. a neural network with 
$N$ three-states
neurons (spins) $s_i (t)\in \{ -1,0,1\}$, $i=1,\ldots ,N$. For each neuron
$s_i$, 
$K$ input sites $j_1 (i),\ldots ,j_K (i)$ are randomly chosen 
among the $N$ 
sites, and $NK$ synaptic interactions $J_{ij}$ are introduced. We 
assume that the synapses are two-states variables 
$J_{ij}\in\{-1,1\}$, randomly and independently sampled with mean $J_0$;
they are not assumed to evolve in time (the case of adapting synapses
is studied in~\cite{cs}).
A parallel deterministic dynamics is assumed for neurons, where the local 
field acting on neuron $s_i$ (the post synaptic potential) is given by
\begin{equation}
h_i(t)=\sum_j J_{ij} s_j(t) \quad ,
\label{eq:lf}
\end{equation}
with the sum taken over the $K$ input neurons. 
We assume a nonmonotonic transfer 
function, depending on the parameter $\theta$~\cite{net1,net2,nn2,net3,net4}:
\begin{equation}
s_i(t+1)=F_\theta \left(h_i (t)\right) \quad ,
\label{eq:nonmon}
\end{equation}
where $F_\theta (x) = \mbox{sign}(x)$ when $|x|<\theta$ and  vanishes 
otherwise.

The dynamics of this model is solved by macroscopic flow equations for
the parameters describing the system.  
Let us  now introduce order parameters for the neurons. The overlap
with pattern $\{\xi\}$ to be retrieved (we choose $\{\xi =1\}$ for simplicity)
is measured by $m(t)=\langle s(t)\rangle$. We stress that the
suppression of the site index $i$ is possible because all averages are 
site-independent. The neuronic activity is given by 
$Q(t)=\langle s^2(t)\rangle$.
The flow equations for $m$ and $Q$ have been obtained in ~\cite{cs}:
\begin{equation}
m(t+1)=\mbox{erf} \left({\mu (t)\over \sqrt{\sigma(t)}}\right) - {1\over 2}
\left [\mbox{erf} \left({\theta+\mu (t)\over \sqrt{\sigma(t)}}\right) - 
\mbox{erf} \left({\theta-\mu (t)\over \sqrt{\sigma (t)}}\right) \right ],
\label{eq:flusso1}
\end{equation}

\begin{equation}
Q(t+1)={1\over 2}
\left [\mbox{erf} \left({\theta+\mu (t)\over \sqrt{\sigma (t)}}\right) + 
\mbox{erf} \left({\theta-\mu (t)\over \sqrt{\sigma (t)}}\right)\right],
\label{eq:flusso2}
\end{equation}
where
\begin{equation}
\mu(t)=K m(t) J_0
\label{eq:limc33}
\end{equation}
and
\begin{equation}
\sigma (t)=K\left(Q(t) - J_0^2 m^2(t)\right)
\label{eq:var}
\end{equation}
are mean and variance, respectively, of the local field acting on neurons at 
time $t$.

Depending on the value of $\theta$ and $J_0$, three kinds of dynamic
behaviour are possible for the network, which lead to a phase 
diagram~\cite{cs}.
Two fixed point ordered phases are present, the {\it ferromagnetic} 
phase (F) characterized by $m>0,Q>0$ and the {\it self-sustained activity} phase
(S) characterized by $m=0,Q>0$. A 
phase without fixed points, corresponding to cyclic or chaotic attractors
and characterized by 
$m_t >0,Q_t >0$, is 
also found; we call it {\it period-doubling} phase (D). We
remark that the phase corresponding to the fixed-point $(m=0,Q=0)$ is
missing in this model. According to the values of the parameters we can get a 
phase or another; in Fig.~1 the bifurcation diagram of $m$ versus
$J_0$, while keeping $\theta =5$ fixed, is shown (for $K=10$).  The S fixed 
point is stable for $J_0 < 0.5$; at $J_0\sim 0.5$ the F fixed point continously 
appears and remains stable until $J_0 \sim 0.69$, where a bifurcation to a 
stable 2-cycle take place. The
bifurcation mechanism is {\it period doubling}, i.e. an eigenvalue of
the Jacobian matrix at the fixed point leaves the unitary disk passing
through $-1$. Increasing $J_0$, successive bifurcations arise; eventually the 
system enters in the chaotic region at $J_0 \sim 0.88$. In the chaotic region 
windows of periodicity intercalate with chaotic attractors, which is a 
well-known feature of dynamical systems with chaotic behaviour. 
We verified that the values of $J_0$ where the successive
bifurcations take place are consistent with Feigenbaum's universality 
law~\cite{feigen}, i.e. 
the length in $J_0$ of the range of stability for an orbit of period $2^n$ 
decreases approximately geometrically with $n$ and the ratio of successive 
range lengths is close to $4.669...$ for large $n$.
We note that, in the phase $D$, the two-dimensional map (3-4)
still possesses the $F$ and $S$ fixed points, but they are unstable.

Let us now consider the strange attractor and its dependence on $J_0$.
For example, in Fig.~2 the strange attractor is shown, for $\theta =5$
and $K =10$, in correspondence with $J_0=0.9$, 0.95 and 0.99 respectively; 
the fixed point F is represented by a star.
In the case $J_0 =0.9$, the attractor is made of two
disconnected components; in the stationary regime successive points  
on the attractor jump from one
component to the other. As $J_0$ grows ($J_0 =0.95$), the attractor
evolves into a more complicated structure, still composed of two
disconnected components. We remark that in these two cases the fixed
point F is not a limit point of the attractor.
At $J_0 =0.99$ the two components of the attractor merge and F 
becomes a limit point of the attractor.

Concerning the Hausdorff dimension of the attractor, we found it to be
close to 0.95 in the three cases described above (the dimension was
estimated by the method described in~\cite{grass}, see also~\cite{eck}).

Let us now discuss the hyperbolicity of the attractor. We remind that in the 
hyperbolic case many interesting properties about the structure and dynamics of 
chaos hold (see, e.g. ~\cite{ott,arrow}). In Fig.3 we have shown finite length 
segments of the stable manifold of the fixed point F (for $J_0 =0.99, \theta = 
5, K=10$); since F is on the attractor, we can argue that the attractor is the 
closure of the unstable manifold of F. We note that the stable and unstable 
manifolds have homoclinic intersections. Moreover Fig.3 displays near 
tangencies between the stable and unstable manifolds; it is reasonable that 
some other segment of the stable manifold will be exactly tangent to the 
unstable manifold. We conclude, therefore, that the attractor is not hyperbolic 
(see~\cite{ott} for a discussion on the non-hyperbolicity of Henon's 
attractor~\cite{henon}, which has similarities with the attractor of the map 
here considered).

We continue our analysis of the dynamical properties of the neural model and we 
turn 
now to the Lyapunov exponents. By evaluating the Jacobian of the map, we find 
it to be area contracting, therefore there is one positive Lyapunov exponent at 
most. We have evaluated  
the first ($\lambda_1$) and the second Lyapunov ($\lambda_2$) exponents by
the method of Ref.~\cite{wolf} and the results are 
displayed in Figs 4a) and 4b): as expected, the second
Lyapunov exponent is always negative. For a given model,
the ratio of the number of free parameters to  the number of 
positive Lyapunov exponents seems to be related, according to a conjecture 
in~\cite{barr}, to the fragility of chaos: if this ratio is greater or equal to 
one a slight change of the parameters can tipically destroy chaos and 
a stable periodic orbit sets in. Since in our case the ratio is $2$ 
(we have two 
parameters, $J_0$ and $\theta$) we conclude that the macroscopic chaos 
of the model should be fragile. 
In Fig.~5 a portion of the parameter
space is depicted; here black pixels correspond to chaotic attractors
and white pixels to stable periodic attractors. One can see that 
extended periodicity windows are present; they are apparently everywhere dense.
The above cited conjecture in~\cite{barr} is based on the idea that periodicity 
windows are constructed around {\it spine loci}, i.e. values of the parameters 
that give rise to superstable orbits. For two dimensional maps, the spine 
locus of cycles with period $p$ is determined by the conditions 
$det M=0$ and $tr M=0$, where $M$ is the Jacobian matrix of the 
$p$-iterated map. Our
map has no critical points (like, e.g., the Henon's map) because the 
determinant of the Jacobian never vanishes; hence the condition $det M=0$ can 
not be strictly satisfied. However, since the map is area contracting,  
$det M \sim 0$ for a periodic orbit with sufficiently high period $p$ (see 
~\cite{barr}): one can therefore 
neglect the condition $det M=0$ and the stability 
requirements reduce to one condition for stability, which according to the 
discussion above is  
$tr M=0$. It follows that the spines are 
of codimension one, i.e. one-dimensional manifolds in the $J_0-\theta$ plane.
In Fig.~6 we have shown finite length segments of the spine loci determined by 
the condition $tr M=0$ with $p=32,64$, and $24$.
Since we do not have an analytic 
treatment of the two-dimensional map, we have implemented the condition $tr 
M=0$ numerically. White areas are periodicity windows which are apparently 
constructed around spine loci; hence 
the conjecture in~\cite{barr} is confirmed. 
We note, in passing, that our findings 
confirm that the behaviour of two-dimensional area-contracting maps is often 
similar to that of one-dimensional maps with critical points~\cite{nn3}.

The concept of robust chaos  as associated with an attractor for 
which the number of positive Lyapunov exponents (in some 
region of parameter space) 
is larger than the number of free (accessible) parameters in the model has been 
also discussed in~\cite{nn4};
recently it has been pointed out that nonmonotonic transfer functions
may lead to robust chaos in time series generated by feed-forward neural
nets~\cite{priel}. Probably fully connected networks with nonmonotonic 
activation function might provide robust chaos; however the analytic analysis 
of these models is difficult since the dynamical theory which describe the 
fully connected Hopfield model~\cite{cool} has not yet been extended to the 
case of nonmonotonic neurons.

\section{Damage spreading.}

A system is said to exhibit damage spreading  if the distance
between two of its replicas, that evolve from slightly different initial
conditions, increase with time (see, e.g.,~\cite{der}). Even though damage 
spreading was first introduced in the context of biologically motivated 
dynamical systems~\cite{kauf}, it has become  an important tool  
to study the influence of initials conditions on the time evolution of 
various physical systems.
In \cite{de}
this phenomenon was studied in diluted networks with monotonic
transfer function. 
The occurrence of damage spreading in 
the Little-Hopfield neural networks, both for fully connected and strongly 
diluted systems,  has been studied in \cite{def}.
Here we generalize this study to the case of diluted
networks with nonmonotonic neurons. Let us consider two replicas of the same 
system having 
initial $t=0$ configurations with the same activity $Q_0$
and overlap $m_0$ but microscopically different 
for a small number of neurons. Subsequently the two replicas evolve,
subject to the same dynamics since their synaptic connections are identical.
The two replicas will have the same macroscopic parameters
$m(t)$ and $Q(t)$ at every
later time $t$, since the trajectories $m(t)$ and $Q(t)$ are obtained
by $m_0$ and $Q_0$ iterating equations (3-4). At microscopic level the 
situation may be different and we therefore study a suitably defined 
distance between the two replicas. Let us call
$h^1 (t)$ and $h^2 (t)$ the local fields acting on $s^1$ and $s^2$, two
corresponding neurons of the replicas located on the same lattice site.
The distance between the local fields is defined by:
\begin{equation}
d(t)= \langle \left( h^1 (t) -h^2 (t) \right )^2 \rangle = 
2 \left( \sigma(t)-\Delta(t)\right ) \quad ,
\label{eq:ds1}
\end{equation}
where
\begin{equation}
\Delta (t)= \langle h^1 (t) h^2 (t) \rangle - \langle h^1 (t)\rangle
\langle h^2 (t)\rangle
\label{eq:ds2}
\end{equation}
is the linear correlation between local fields at time $t$. In the limit
$N\to\infty$ with $K$ large and finite, $h^1$ and $h^2$ can be treated
as gaussian variables with probability density:

\begin{equation}
P_t(h^1,h^2)={1 \over C} \exp
\left\{ -{1\over 2} 
\left[
{\sigma \over {\sigma^2 - \Delta^2}}
\left( (h^1 -\mu )^2 +(h^2 -\mu)^2 \right)
-{2\Delta\over\sigma^2-\Delta^2}(h^1-\mu)(h^2 -\mu) 
\right]
\right\} \quad ,
\label{ds3}
\end{equation}
where $C$ is a normalization factor and $\sigma$, $\mu$, $\Delta$ are
implicitly dependent on the time $t$. The time evolution law for $\Delta
(t)$ is given by:

\begin{equation}
\Delta (t+1) = K\left(\langle s^1 (t+1) s^2 (t+1)\rangle -J_0^2
m^2(t+1)\right).
\label{ds4}
\end{equation}
The average of the product of corresponding neurons in the two replicas
is evaluated as follows:

\begin{equation}
\langle s^1 (t+1) s^2 (t+1)\rangle = \int dh^1 \int dh^2 P_t (h^1, h^2 )
F_\theta (h^1) F_\theta (h^2);
\label{ds5}
\end{equation}
the evaluation of the integral on the r.h.s. of (\ref{ds5}) is straightforward 
and leads to the time evolution law for $\Delta (t)$, which  can be
written as follows:

\begin{equation}
\Delta (t+1) = K\left ( \int_{-{\mu \over \sqrt{\sigma}}}^{\theta - \mu
\over \sqrt{\sigma}} Dz I(z) -\int_{-{\theta + \mu
\over \sqrt{\sigma}}}^{-{\mu \over \sqrt{\sigma}}} Dz I(z) - J_0^2 m^2
(t+1)\right) \quad ,
\label{ds6}
\end{equation}
where $Dz=e^{-{1\over 2}z^2} {dz\over \sqrt{2\pi}}$ is the gaussian
measure and

\begin{equation}
I(z) = \mbox{erf}(A) + {1\over 2}[\mbox{erf}(B) - \mbox{erf}(C)],
\label{ds7}
\end{equation}
with

\begin{equation}
A={\mu\sqrt{\sigma}\over \sqrt{\sigma^2 - \Delta^2}} +
{\Delta z \over \sqrt{\sigma^2 - \Delta^2}}\;\;\;,
\label{ds8}
\end{equation}

\begin{equation}
B={(\theta-\mu)\sqrt{\sigma}\over \sqrt{\sigma^2 - \Delta^2}} -
{\Delta z \over \sqrt{\sigma^2 - \Delta^2}}\;\;\;,
\label{ds9}                                                 
\end{equation}

\begin{equation}
C={(\theta+\mu)\sqrt{\sigma}\over \sqrt{\sigma^2 - \Delta^2}} +
{\Delta z \over \sqrt{\sigma^2 - \Delta^2}}\;\;\;.
\label{ds10}                                                 
\end{equation}

Equation (\ref{ds6}), together with (3-4), solves the time evolution of 
$\Delta (t)$. We
remark that $\Delta (t) = \sigma(t)$ (i.e. $d(t)=0$) is a fixed point
for eq.(\ref{ds6}). The possible occurrence of damage spreading can be now seen
to be equivalent to 
the instability of the fixed point $\Delta=\sigma$. We have studied this 
problem numerically.
We find that damage spreading occurs for
any choice of the parameters $J_0$, $\theta$, $K$. It follows that,
from a microscopic point of view, the motion of the system is always to be
considered chaotic even though at the macroscopic level it can exhibit 
different behaviours (fixed point, periodicity, chaos). In fig.~7 we depict 
the stationary regime of $d(t)$ in
the case of periodic macroscopic dynamics and chaotic dynamics; the initial 
distance was $d(0)=10^{-5}$. The macroscopic 
behaviour can be seen in the lower part of the figures (the overlap trajectory 
$m(t)$); it is a cycle with period 4 in case (a) and it is chaotic in case (b).
Correspondingly the distance $d(t)$ is always greater than zero (i.e. damage 
spreading occurs) and has period 4 in case (a) whereas it shows chaotic 
behaviour in case (b). 

We remark that in Ref.~\cite{cs} it has been shown that the presence of 
adapting synapses in these networks leads to reduction of macroscopic dynamics. 
The adapting system self-regulates its synaptic configuration, by its own 
dynamics, so as to escape from chaotic regions: in the stationary regime, the 
mean of synapses $J(t)$ remains practically constant (equal to the stationary 
value $J_{stat}$) and the system settles in periodic macroscopic orbits. Since 
we found damage spreading for all fixed values of $J_0$, it follows that the 
adapting system also displays damage spreading in the stationary regime. 
In other words, the adaptiveness of synapses 
should not remove damage spreading although it reduces the macroscopic dynamics.
However damage spreading may be suppressed
if the neuron updating rule becomes stochastic by  a 
proper amount of noise (the two replicas being subject to the same noise, 
see~\cite{molg,der,def}).

\section{Conclusions.}

In this paper we have studied a diluted neural network with nonmonotonic
activation function whose macroscopic dynamics is given by a
two-dimensional map. Some properties of this non-linear map have been studied:
the structure and non-hyperbolicity of the strange attractor; in particular we 
have analyzed 
the fragility of the chaos and we have shown the validity of a recently 
discussed conjecture, i.e. periodicity windows are constructed around spine 
loci. Finally we have studied the time evolution of
the distance between two replicas of the model which evolve subject to
the same synaptic configuration. We have found that the two replicas never
become identical, and the system exhibits damage spreading for any
choice of the parameters. 
In the stationary regime the distance between the two replicas does not
vanish and the trajectory $d(t)$ behaves in agreement with the macroscopic 
dynamics: it is periodic (chaotic) if $m(t)$ is periodic (chaotic). 

\section*{Ackowledgements}

The authors gratefully thank L. Angelini, G. Gonnella,
M. Pellicoro and M. Villani for 
useful discussions.

\newpage
\noindent\Large\textbf{Figure Captions}
\normalsize
\vspace{1.0cm}
\begin{description}
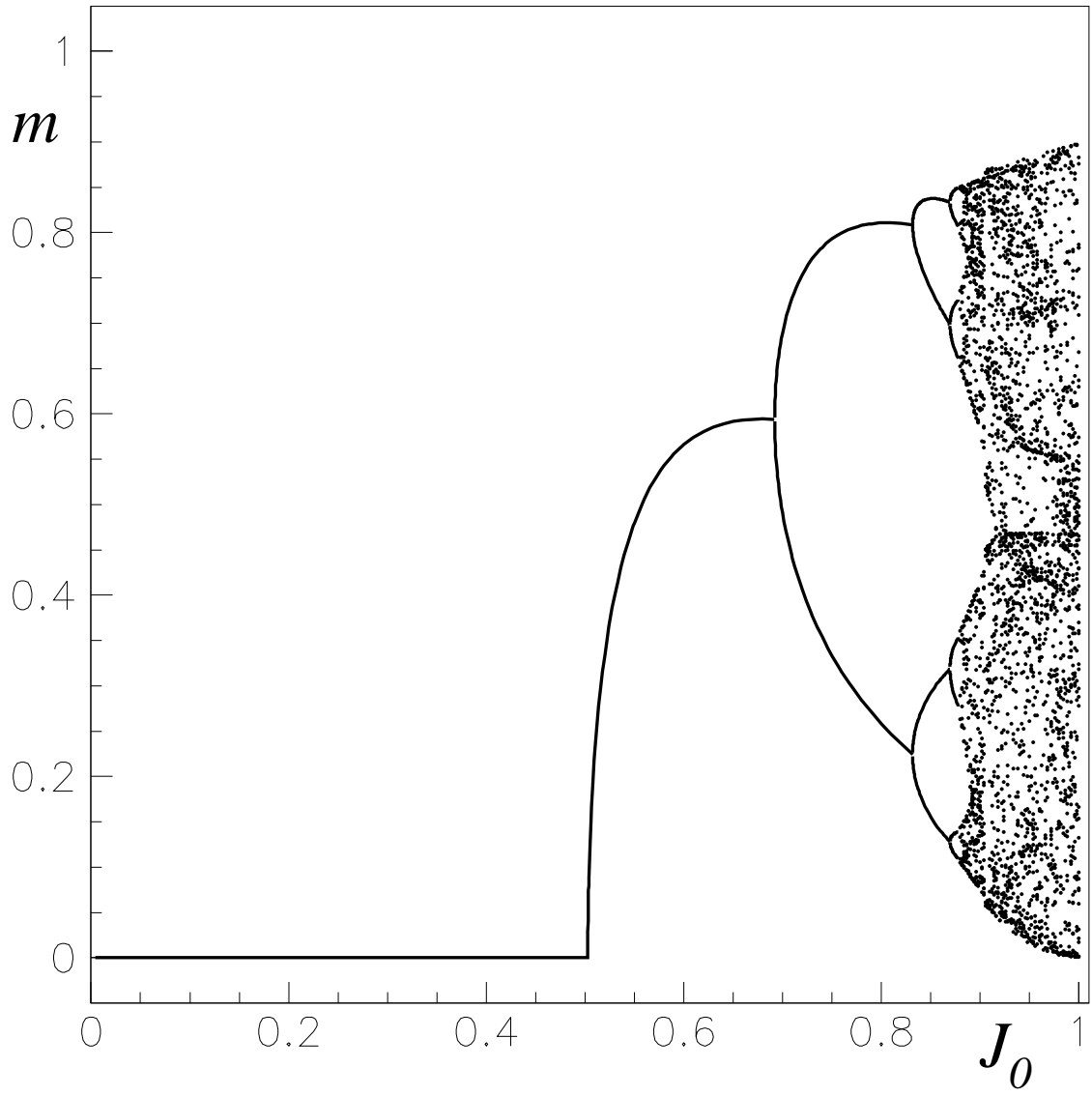
\item{Figure 1}: 
Bifurcation map of $m$ versus $J_0$ in the case $K=10$ and $\theta=5$.

\item{Figure 2}: 
The strange attractor of map (3-4), corresponding to $K=10$, $\theta=5$
and $J_0 = 0.9$ (a), $0.95$ (b), $0.99$ (c). The star represents the fixed 
point F.

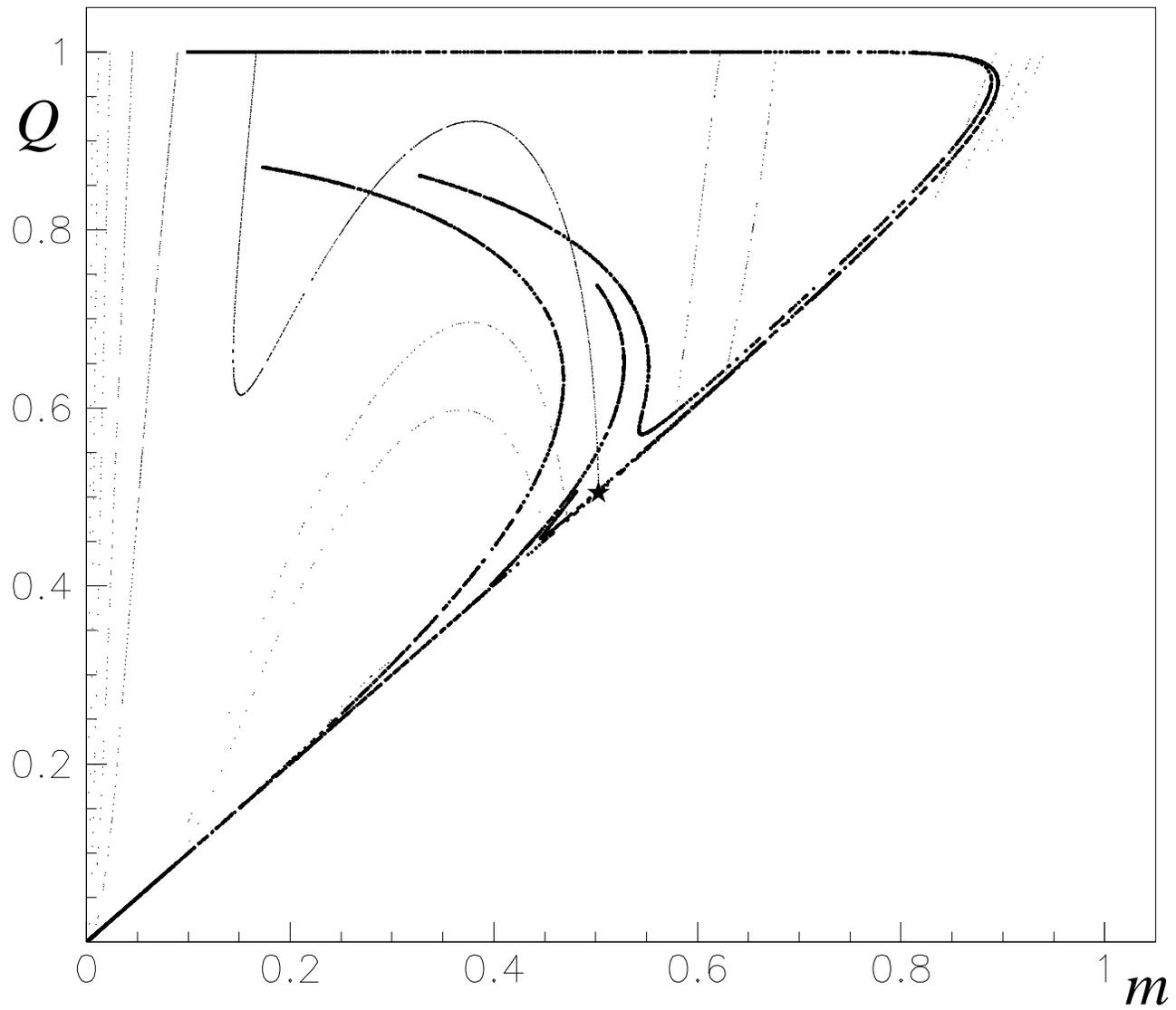
\item{Figure 3}: 
The stable and unstable manifolds of the  fixed point F (represented by the 
star), for $J_0 = 0.99, K=10,\theta=5$.

\item{Figure 4}: 
First (a) and second (b) Lyapunov exponents versus $\theta$,
corresponding to $K=10$ and $J_0=0.9$.

\item{Figure 5}: 
A portion of the parameter space: 
$(\theta,J_0)\in [5,5.12]\times [0.88,0.9]$.
Black pixels correspond to chaotic bahaviour whereas white pixels correspond to
periodicity.

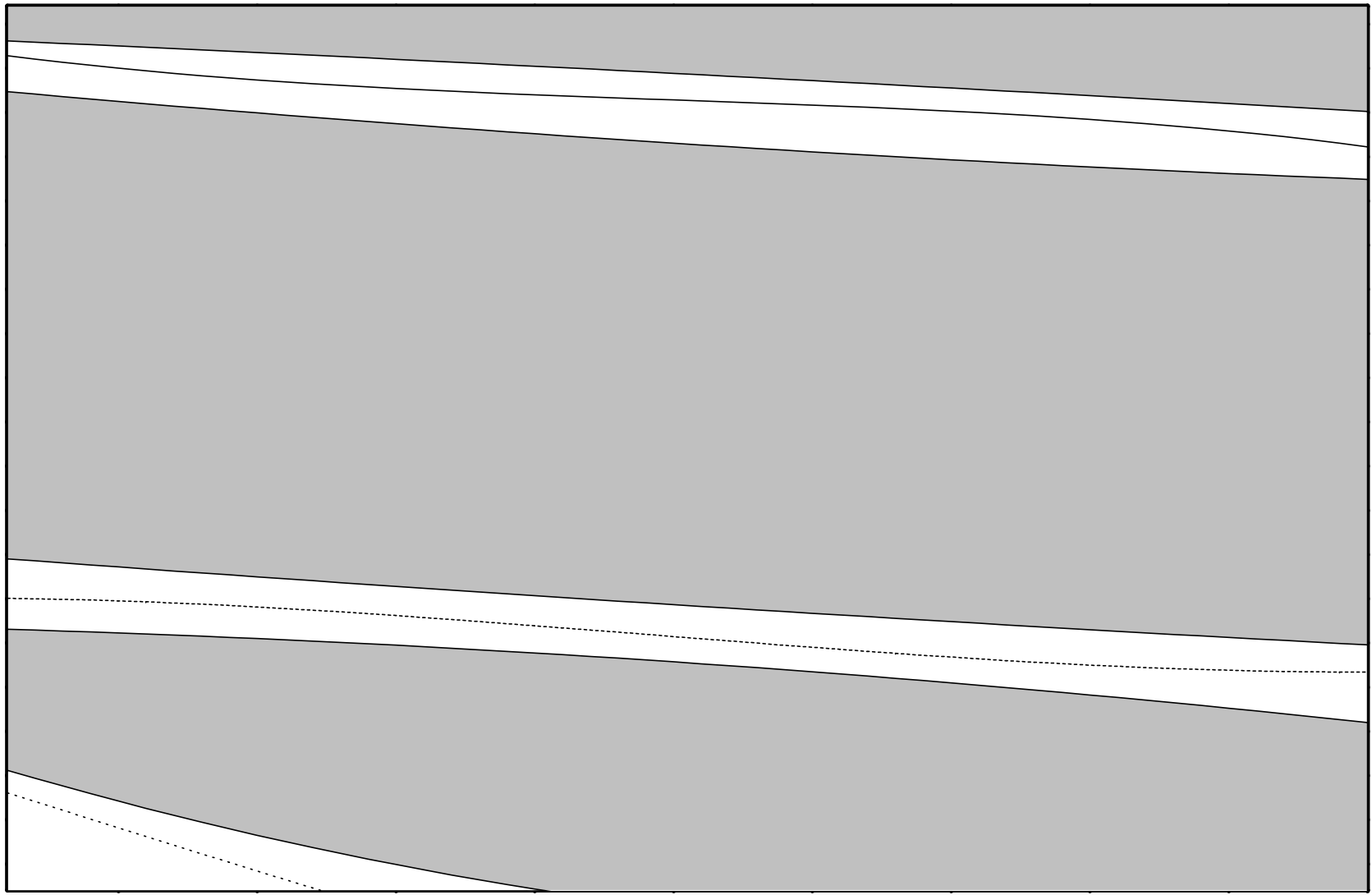
\item{Figure 6}:
A portion of parameter space:
$(\theta,J_0)\in [5,5.015]\times [0.88,0.8825]$.
White areas correspond to periodic behaviour. The solid line represents the 
spine locus with $p=24$, the dashed line is the spine locus with $p=64$, the 
dotted line is the spine with $p=32$. These curves are obtained numerically by 
interpolation of a finite set of points characterized by the condition $tr M=0$
. Grey areas correspond to chaotic behaviour; they contain infinite periodicity 
windows not displayed here.

\item{Figure 7}: 
The stationary regime of the distance between two configurations having
initial distance $d(0)=10^{-5}$. Squares correspond to the distance: $y=d(t)$,
while triangles  represent the overlap: $y=m(t)$. 
The parameter values are $K=10,
\theta=5, J_0=0.85$ (a) and $K=10, \theta=5, J_0=0.95$ (b). 

\end{description}
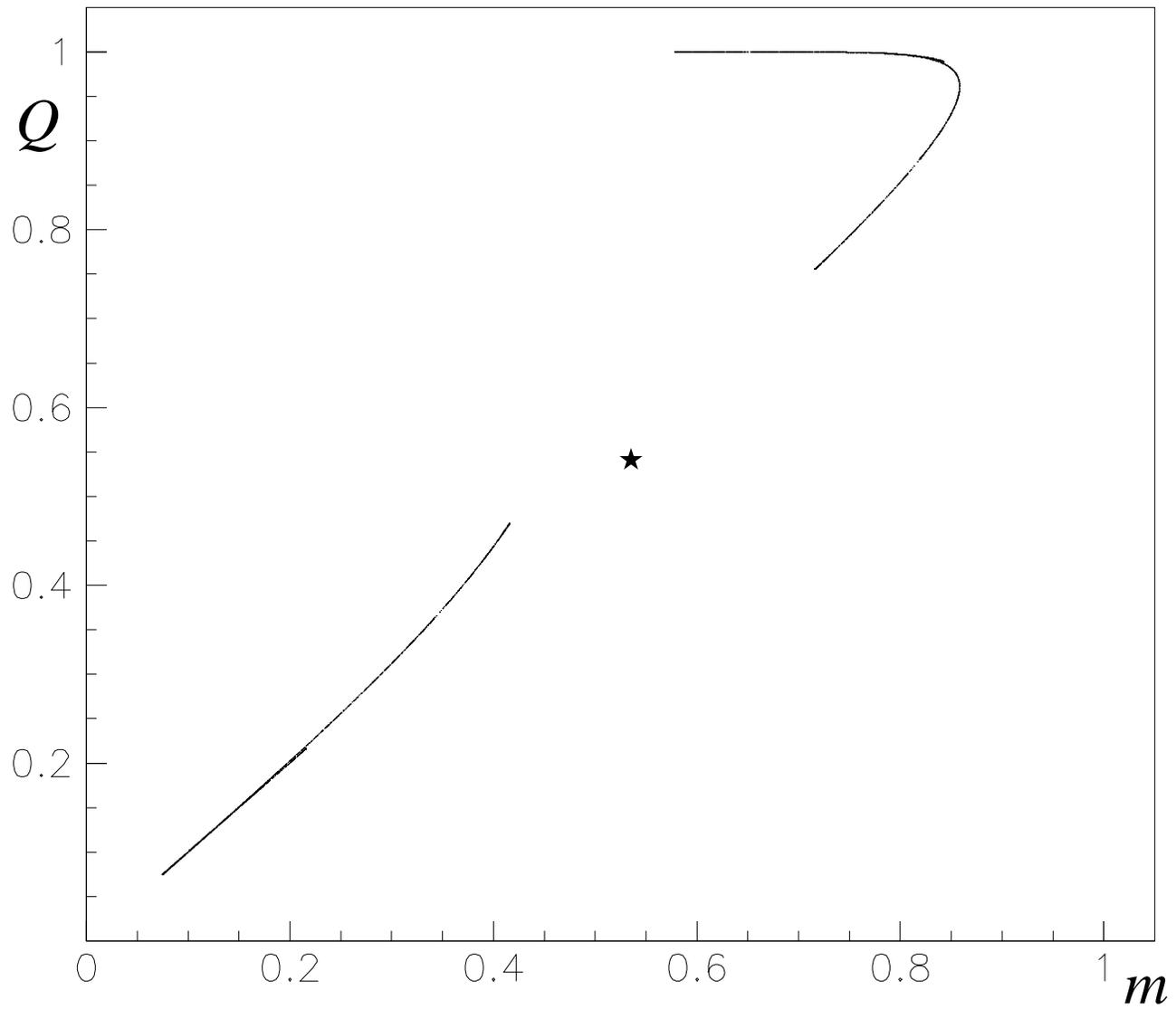
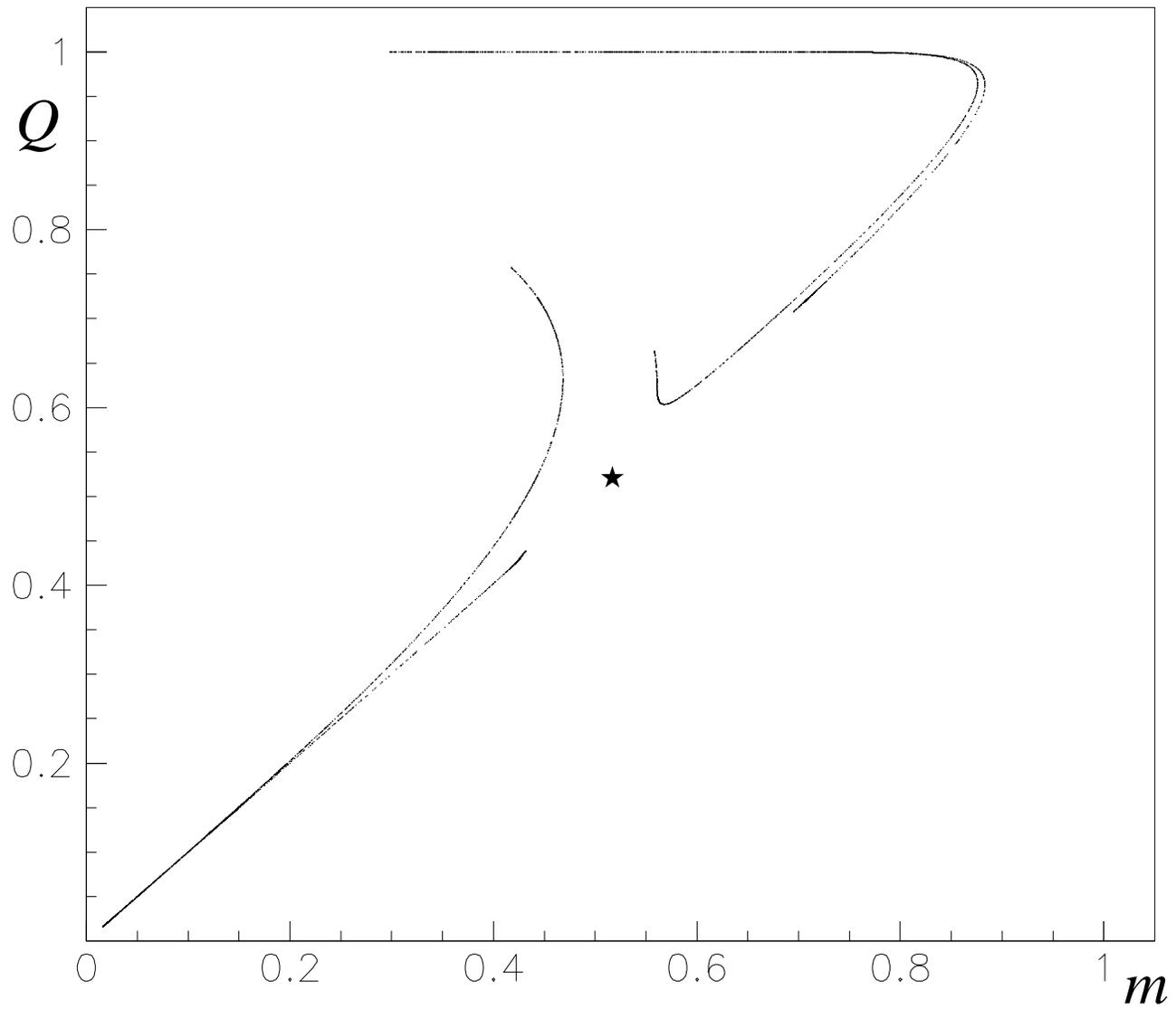
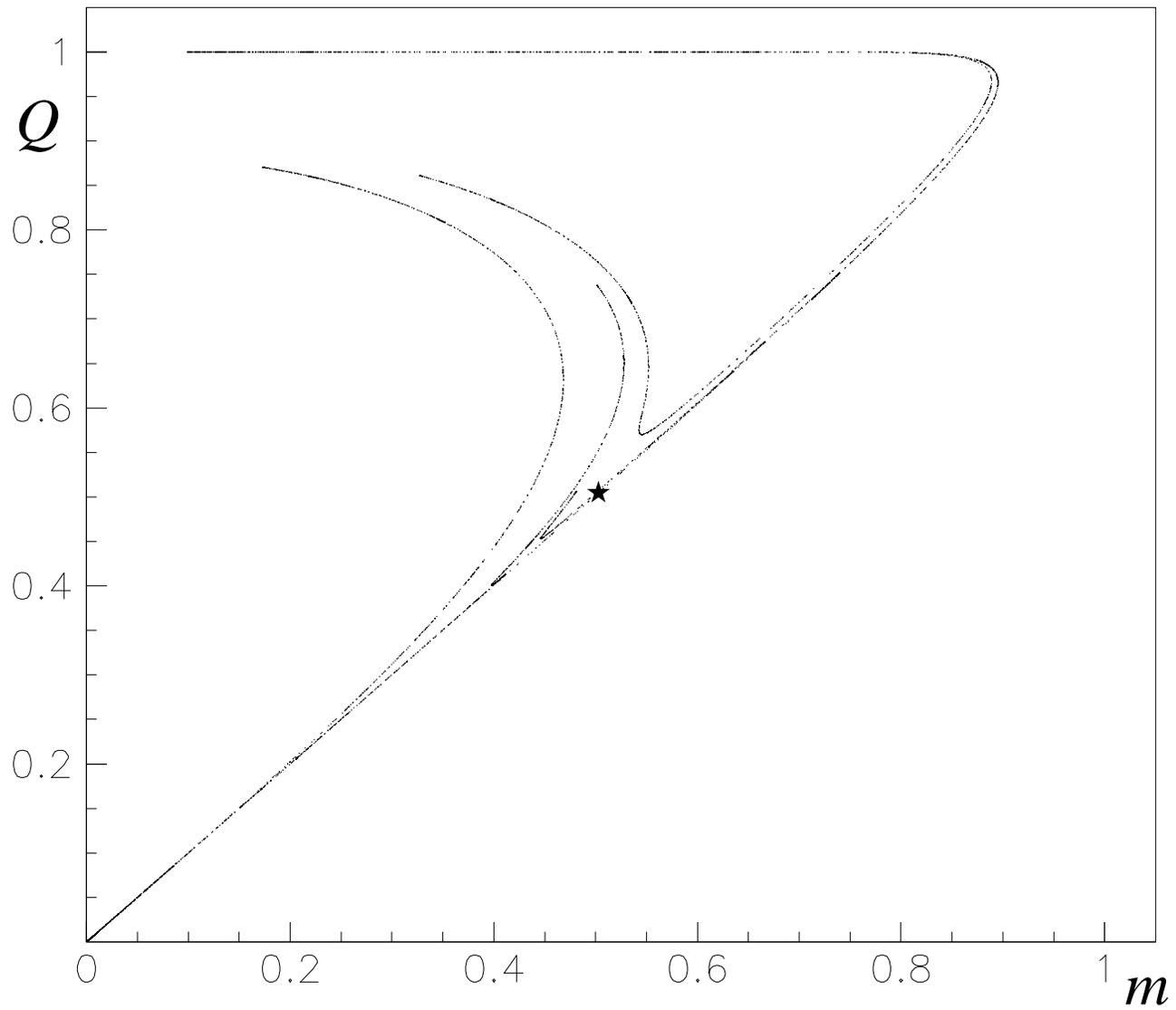
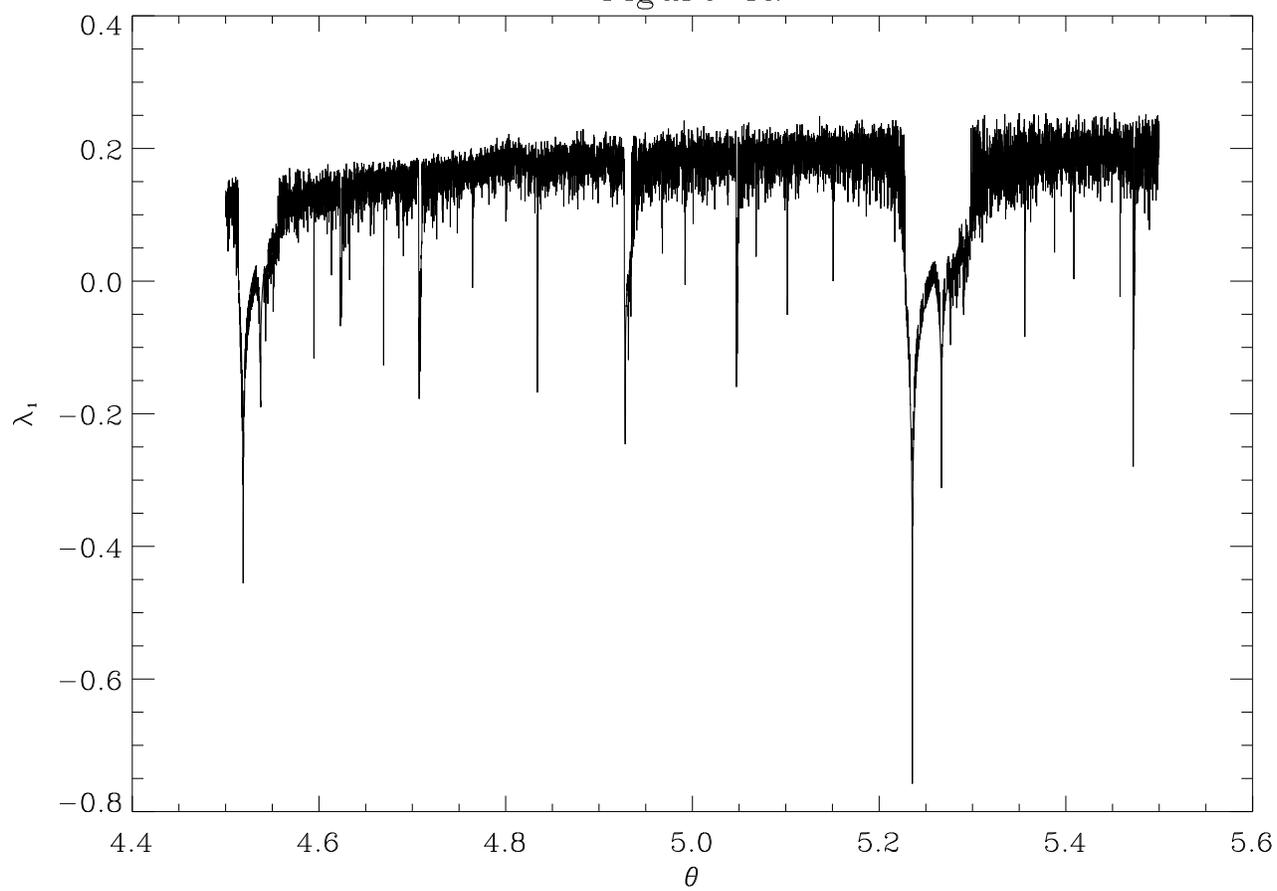
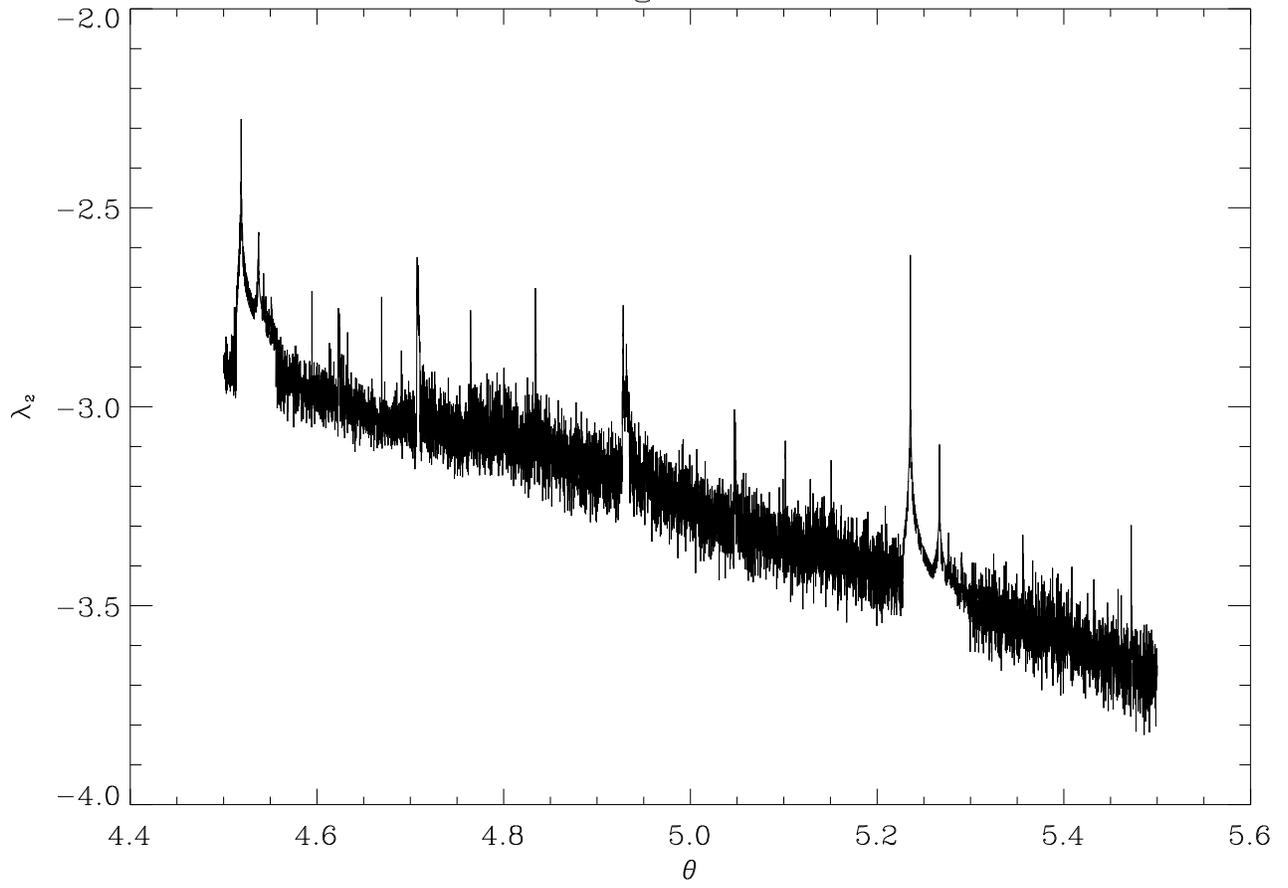
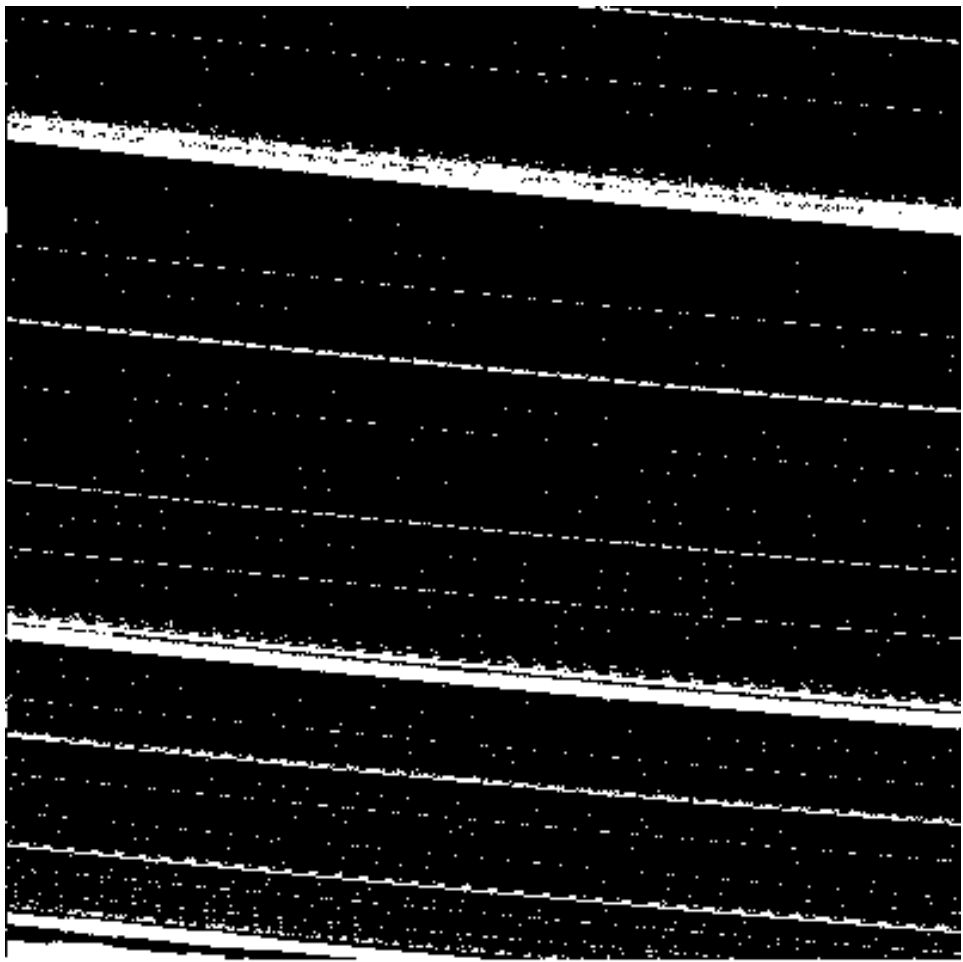
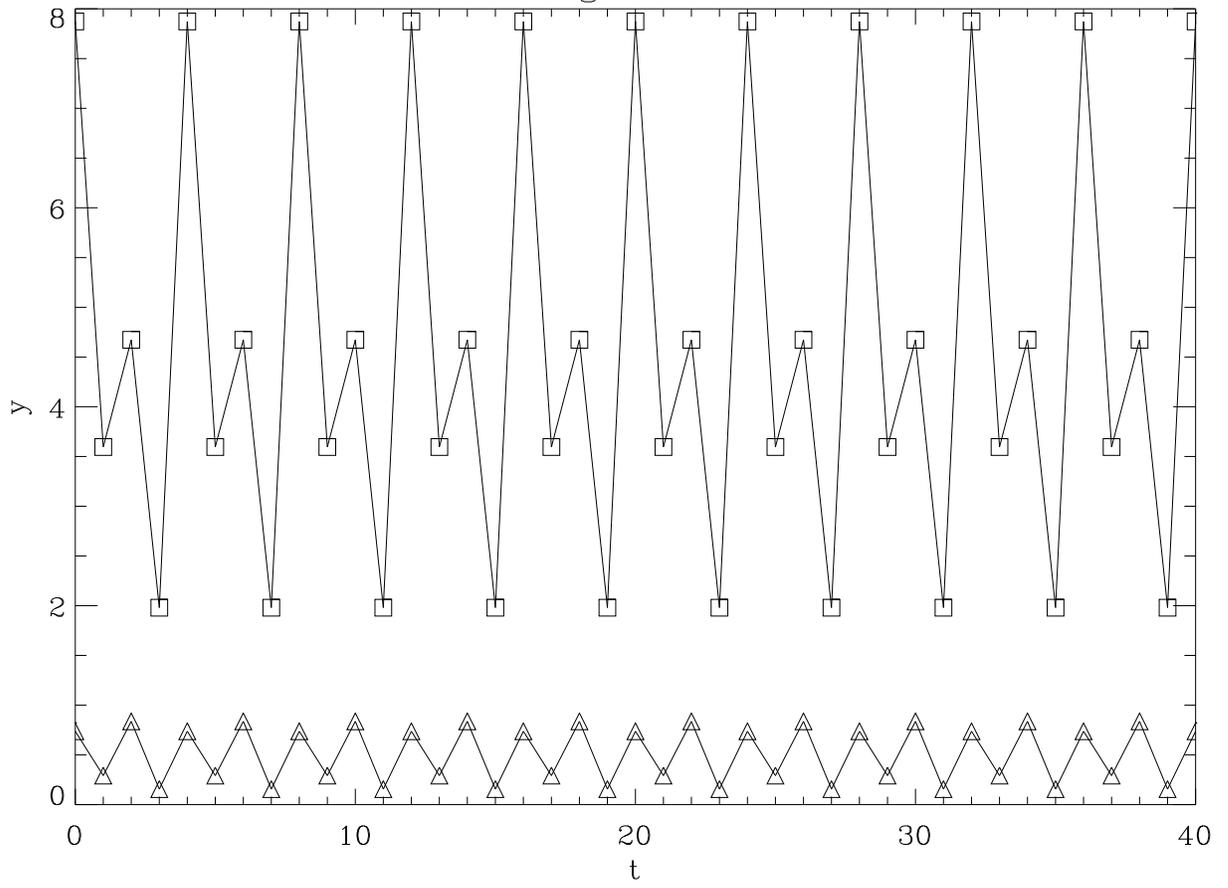
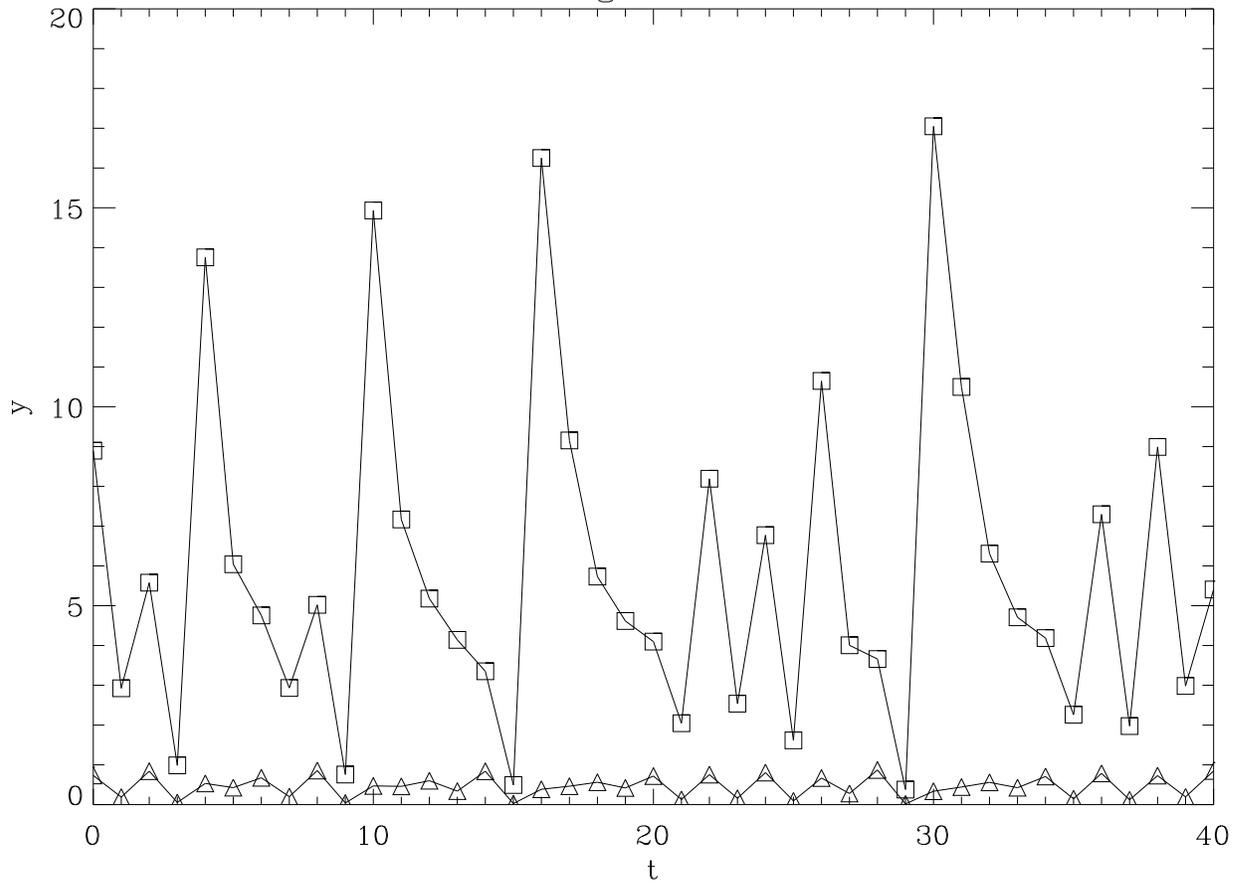
\end{document}